\documentclass[sigconf]{acmart}

\usepackage{amsmath}
\usepackage{amsfonts}
\usepackage{graphicx}
\usepackage{hyperref}
\usepackage{booktabs}
\usepackage{multirow}
\usepackage{tabularx} 
\usepackage{booktabs} 

\usepackage{url}
\usepackage[dvipsnames]{xcolor}
\usepackage{enumitem}
\usepackage{algorithm}
\usepackage{algpseudocode}
\usepackage{pifont}

\usepackage[many]{tcolorbox} 
\usepackage{fontawesome5} 
\usepackage{colortbl}
\usepackage{threeparttable}
\usepackage{makecell}
\definecolor{checkgreen}{rgb}{0,0.6,0}
\definecolor{lightgreen}{rgb}{0.9,1,0.9}
\newtcolorbox{boxK}{
    top=2pt,
    bottom=2pt,
    left=2pt,
    right=2pt,
    boxrule = 0pt,
    toprule = 0pt, 
}

\title{ClassEval-Pro: A Cross-Domain Benchmark for Class-Level Code Generation}

\author{Yeheng Chen}
\authornote{These authors contributed equally.}
\affiliation{%
  \institution{Shanghai Jiao Tong University}
  \country{China}
}

\author{Chaoxiang Xie}
\authornotemark[1]
\affiliation{%
  \institution{Hohai University}
  \country{China}
}

\author{Yuling Shi}
\affiliation{%
  \institution{Shanghai Jiao Tong University}
  \country{China}
}

\author{Wenhao Zeng}
\affiliation{%
  \institution{Shanghai Jiao Tong University}
  \country{China}
}

\author{Yongpan Wang}
\affiliation{%
  \institution{Shanghai Jiao Tong University}
  \country{China}
}

\author{Hongyu Zhang}
\affiliation{%
  \institution{Chongqing University}
  \country{China}
}

\author{Xiaodong Gu}
\authornote{Corresponding author.}
\affiliation{%
  \institution{Shanghai Jiao Tong University}
  \country{China}
}
\email{xiaodong.gu@sjtu.edu.cn}
\begin{document}

\begin{abstract}
LLMs have achieved strong results on both function-level code synthesis and repository-level code modification, yet a capability that falls between these two extremes---\textit{compositional code creation}, i.e., building a complete, internally structured class from a specification---remains underserved. Current evaluations are either confined to isolated functions or rely on manually curated class-level tasks that are expensive to scale and increasingly susceptible to data contamination. We introduce ClassEval-Pro, a benchmark of 300 class-level tasks spanning 11 domains, constructed through an automated three-stage pipeline that combines complexity enhancement, cross-domain class composition, and integration of real-world GitHub code contributed after January 2025. Every task is validated by an LLM Judge Ensemble and must pass test suites with over 90\% line coverage. We evaluate five frontier LLMs under five generation strategies. The best model achieves only 45.6\% class-level Pass@1, with a 17.7-point gap between the strongest and weakest models, confirming the benchmark's discriminative power. Strategy choice strongly interacts with model capability: structured approaches such as bottom-up improve weaker models by up to 9.4 percentage points, while compositional generation collapses to as low as 1.3\%. Error analysis over 500 manually annotated failures reveals that logic errors (56.2\%) and dependency errors (38.0\%) dominate, identifying cross-method coordination as the core bottleneck.
\end{abstract}

\maketitle

\section{Introduction}
LLMs now excel at two distinct modes of code generation. On function-level benchmarks, frontier models surpass 90\% Pass@1 on HumanEval~\cite{chen2021} and MBPP~\cite{austin2021}, approaching human-level accuracy on isolated programming tasks. On repository-level benchmarks, coding agents resolve over 75\% of real-world GitHub issues on SWE-bench Verified~\cite{jimenez2024swebench}. Yet real software development also demands a capability that falls between these two extremes: building a complete, internally structured class from a specification---coordinating multiple methods, managing shared state, and integrating logic across domains~\cite{hou2024largelanguagemodelssoftware,2011Portfolio}. Current function-level benchmarks (HumanEval, MBPP, BigCodeBench~\cite{zhuo2025bigcodebench}) test isolated logic synthesis, while repository-level benchmarks (SWE-bench, DevEval~\cite{li2024dev}) test code modification within existing codebases; neither evaluates whether models can architect a coherent multi-method artifact from scratch. ClassEval~\cite{du2023classeval} pioneered class-level evaluation, yet its reliance on manual curation (500 person-hours for 100 tasks across 7 domains) makes it difficult to scale, and its 2023-vintage data is increasingly exposed to training-set contamination~\cite{rahman2025,fang2025lastingbench}. We refer to this underserved evaluation dimension as \textit{compositional code creation}.

To address this gap, we introduce ClassEval-Pro, a benchmark of 300 class-level tasks spanning 11 domains, designed to evaluate compositional code creation. Three design choices distinguish it from prior work. (1) \textit{Cross-domain composition}---tasks require integrating heterogeneous domain logic (e.g., configuration management + financial computation) into a single coherent class, testing compositional reasoning that single-domain tasks cannot capture. (2) \textit{Automated, contamination-resistant construction}---an LLM Judge Ensemble pipeline~\cite{gu2025surveyllmasajudge} replaces manual curation, sourcing from GitHub repositories contributed after January 2025~\cite{zhu2024domain} and employing multi-model collaboration for skeleton generation, test alignment, and reference implementation. (3) \textit{Strict validation}---every task is verified against automatically generated test suites that achieve over 90\% line coverage, ensuring that both the specification and the reference solution are functionally correct.

We evaluate five frontier LLMs---GPT-5.1, Gemini-2.5-Pro, Qwen3-480B, Qwen3-30B, and Kimi-K2---under five generation strategies (holistic, incremental, compositional, top-down, and bottom-up) and manually annotate 500 failed cases. Class-level Pass@1 ranges from 27.9\% to 45.6\% (17.7-point gap), far wider than on function-level benchmarks, confirming that compositional tasks discriminate frontier models effectively. Strategy choice strongly interacts with model capability: structured decomposition (bottom-up) improves weaker models by up to 9.4 percentage points, while compositional generation collapses to as low as 1.3\%. Error analysis reveals that logic errors (56.2\%) and dependency errors (38.0\%) dominate failures, identifying cross-method coordination---not surface-level code formation---as the core bottleneck in class-level code generation.

Our major contributions are as follows:
\begin{itemize}[left=0pt]
    \item We introduce \textbf{ClassEval-Pro}, a benchmark of 300 class-level tasks across 11 domains for evaluating \textit{compositional code creation}. It moves beyond isolated function synthesis by requiring models to generate coherent multi-method classes that integrate shared state, inter-method dependencies, interface constraints, and domain-specific logic.

    \item We develop an automated and contamination-resistant construction pipeline for scalable class-level benchmark generation. The pipeline integrates complexity enhancement, cross-domain class composition, post-2025 GitHub code, and LLM Judge Ensembles to ensure task quality, feasibility, and reduced exposure to pretraining contamination.

    \item We present diagnostic findings that characterise the strengths and limitations of current LLMs on class-level code generation. Through strategy--model interaction analysis and a fine-grained error taxonomy, we identify recurring bottlenecks that are not visible from pass/fail metrics alone, providing guidance for future model and agent design.
\end{itemize}

\section{Related Work}

\subsection{Large Language Models for Code Generation}
Large Language Models (LLMs) have significantly advanced code generation~\citep{zhuo2025bigcodebench,shi2024between}, repair~\citep{shi2024code,chen2025swe}, translation~\citep{khan2024xcodeeval,hu2025flowmaltransunsupervisedbinarycode}, and reasoning~\citep{gu2024cruxeval,zeng2025pruning}. This progress is driven by both open-weight models (e.g., DeepSeek-Coder~\citep{guo2024deepseek}, Qwen3-Coder~\citep{yang2025qwen3}) and proprietary models (e.g., GPT-5~\citep{singh2025openaigpt5card}, Gemini~3~\citep{google2026gemini3}). Recent work further extends LLM capabilities to code understanding through vision--language models~\citep{shi2026codeocr} and long-context compression for code~\citep{shi2025longcodezip}, as well as vulnerability detection~\citep{hu2026zeroshotvulnerabilitydetectionlowresource}. Despite these advances, current benchmarks predominantly evaluate atomic function-level tasks or repository-level modifications. Despite object-oriented programming dominating real-world software~\citep{2011Portfolio}, evaluations for compositional class-level creation remain limited.

Recent studies further extend LLM-based code generation from isolated functions to larger software units. Repository-level code generation requires models to reason over dependencies across functions, classes, and modules, and context inlining has been proposed to expose both upstream and downstream dependency information during generation~\citep{hu2026context}. Similarly, project-level code translation studies show that skeleton-guided generation can help preserve compilability and inter-module dependencies when translating complete C projects to Rust~\citep{wang2026evoc2rust}. These works motivate evaluations that stress dependency management and structural consistency beyond isolated function synthesis.

\subsection{Benchmarks for Code Generation}

Early code generation benchmarks target function-level synthesis (e.g., HumanEval~\citep{chen2021}, MBPP~\citep{austin2021}). While subsequent works enhance evaluation rigor~\citep{liu2023code}, API diversity~\citep{zhuo2025bigcodebench}, and data freshness~\citep{jain2024livecodebench}, they fundamentally fail to capture the structural complexity of real-world software~\citep{hou2024largelanguagemodelssoftware}.

To address this, recent evaluations diverge into two trajectories: repository-level modification and compositional generation. Modification benchmarks evaluate issue resolution within existing codebases (e.g., SWE-bench~\citep{jimenez2024swebench}, DevEval~\citep{li2024dev}) or utilize repository context~\citep{zhang2023repo}. More recent repository- and project-level benchmarks further extend this direction by considering refactoring tasks, agentic project completion, and realistic industrial programming scenarios~\citep{xu2026swerefactor,liu2025projecteval,industrycode2026}. These benchmarks better reflect real software engineering workflows, increasingly supported by agentic techniques such as multi-agent debate~\citep{li2025swe}, skill-based optimization~\citep{wang2026effiskill}, and adaptive context pruning~\citep{wang2026swe}. Complementary benchmarks have also been proposed for repository-level question answering~\citep{peng2025swe}. However, they primarily evaluate modification, agent interaction, or repository-context use rather than from-scratch class construction.

Generation benchmarks, conversely, target structural creation, such as context-dependent units (CoderEval~\citep{Zhang_2024}) or entire classes (ClassEval~\citep{du2023classeval}). Recent class-level studies further highlight the gap between synthetic class tasks and real-world class implementations, showing that real-source classes introduce substantially more dependency and context challenges~\citep{rahman2025realclass,rahman2025openclassgen}.

However, reliance on manual curation restricts existing class-level benchmarks in scalability and domain coverage~\citep{du2023classeval,rahman2025}. As shown in Table~\ref{tab:benchmark_comparison}, while several existing benchmarks incorporate \textit{multi-domain evaluation}, none support the automated construction of from-scratch, real-source tasks at the class level. ClassEval-Pro bridges this gap by uniquely integrating an auto-construct pipeline with multi-domain, real-source grounding for from-scratch generation.

\begin{table}[t]
\centering
\definecolor{checkgreen}{RGB}{34,139,34}
\definecolor{crossred}{RGB}{204,0,0}
\definecolor{lightgreen}{RGB}{230,250,230}
\definecolor{sectiongrey}{gray}{0.95}
\newcommand{\ch}{\textcolor{checkgreen}{$\checkmark$}}
\newcommand{\xr}{\textcolor{crossred}{$\times$}}

\caption{Comparison of code generation benchmarks.}
\label{tab:benchmark_comparison}

\resizebox{\columnwidth}{!}{%
\begin{tabular}{lcccccc}
\toprule
\textbf{Benchmark} & \textbf{Task} & \makecell{\textbf{Avg.}\\\textbf{LOC}} & \makecell{\textbf{Cross}\\\textbf{Domain}} & \makecell{\textbf{Auto}\\\textbf{Construct}} & \makecell{\textbf{Real}\\\textbf{Source}} & \makecell{\textbf{From}\\\textbf{Scratch}} \\
\midrule

\rowcolor{sectiongrey}
\multicolumn{7}{l}{\textit{Function-level generation}} \\
HumanEval & 164 & 11.5 & \xr & \xr & \xr & \ch \\
MBPP & 974 & 6.8 & \xr & \xr & \xr & \ch \\
APPS & 5000 & 21.4 & \xr & \xr & \ch & \ch \\
HumanEval+ & 164 & 11.5 & \xr & \xr & \xr & \ch \\
LCBench & 1055 & -- & \xr & \ch & \ch & \ch \\
\midrule

\rowcolor{sectiongrey}
\multicolumn{7}{l}{\textit{Repository-level modification}} \\
SWE-bench & 2294 & -- & \ch & \ch & \ch & \xr \\
DevEval & 1874 & -- & \ch & \xr & \ch & \xr \\
\midrule

\rowcolor{sectiongrey}
\multicolumn{7}{l}{\textit{Class-level creation}} \\
ClassEval & 100 & 45.7 & \ch & \xr & \xr & \ch \\
CoderEval & 230 & 30.0 & \ch & \xr & \ch & \ch \\

\rowcolor{lightgreen}
\textbf{ClassEval-Pro} & \textbf{300} & \textbf{113} & \ch & \ch & \ch & \ch \\
\bottomrule
\end{tabular}%
}
\vspace{-0.6cm}
\end{table}

\subsection{Dataset Augmentation}
Traditional data augmentation relies on semantic-preserving transformations~\citep{sennrich2016,wei2019} to improve generalization. Recently, LLM-driven synthesis has emerged as a scalable alternative, utilizing seed-based bootstrapping (e.g., Self-Instruct~\citep{wang2023selfinstruct}) and iterative complexity escalation (e.g., Evol-Instruct~\citep{xu2025wizardlm}). 
However, these methods primarily target training data generation rather than benchmark curation. Repurposing them for evaluation introduces strict requirements: generated tasks must guarantee functional correctness, align with executable test suites, and prevent training data contamination. Existing frameworks fail to systematically enforce these constraints for compositional class-level generation. ClassEval-Pro bridges this by implementing a rigorous auto-construct pipeline, which integrates LLM-based skeleton generation, multi-model peer validation, and strict line-coverage filtering to synthesize high-quality, multi-domain evaluation tasks.

\section{Benchmark Construction}

This section introduces ClassEval-Pro, a benchmark of 300 multi-domain class-level code generation tasks spanning 11 domains.

\subsection{Pipeline Overview}
\label{sec:pipeline}
\begin{figure}
    \includegraphics[width=\linewidth]{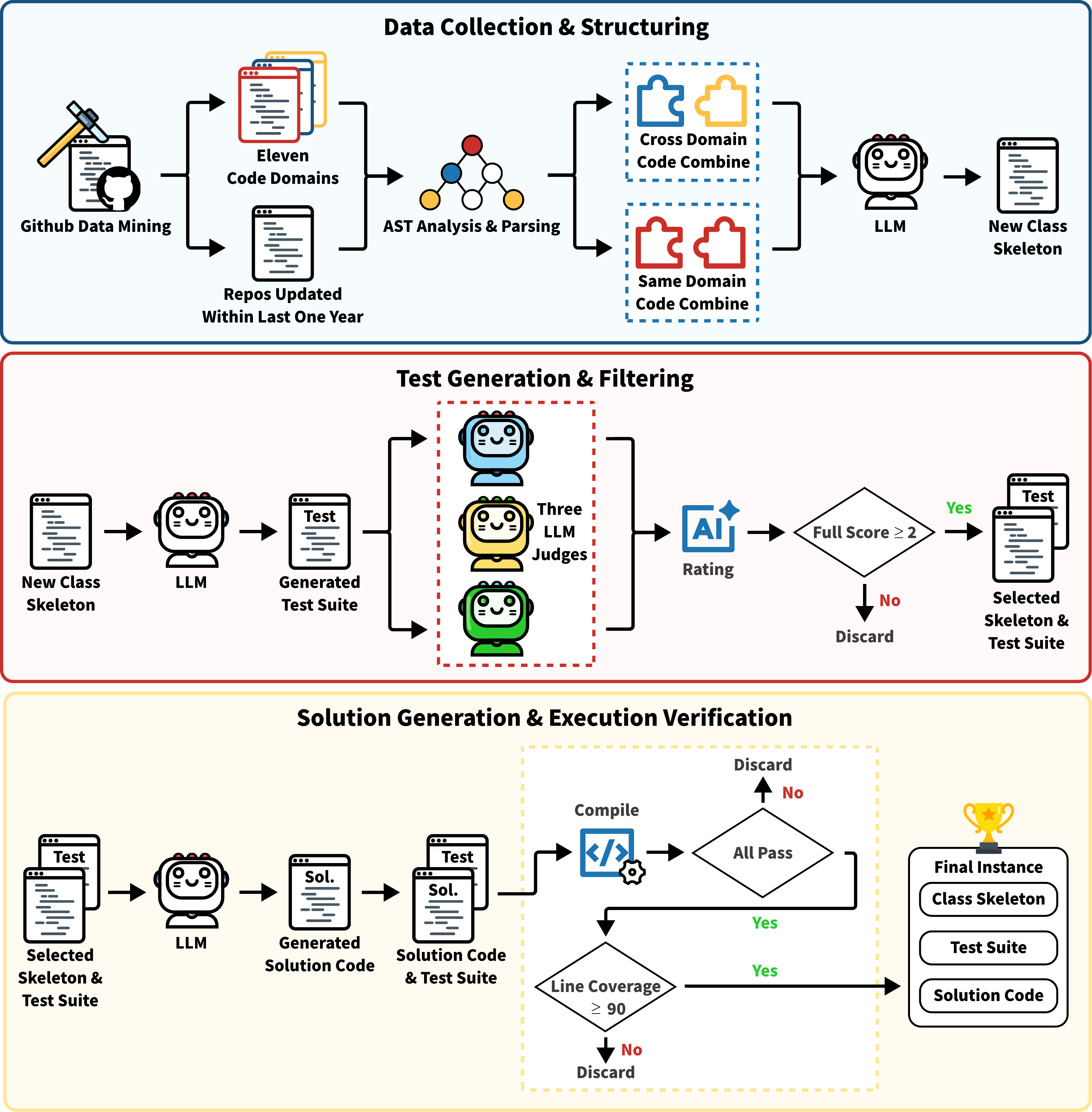}
    \vspace{-0.2cm}
    \caption{Overview of the benchmark construction pipeline.}
    \vspace{-0.7cm}
    \label{fig:Overview}
\end{figure}

To mitigate data contamination~\citep{jain2024livecodebench} and the manual overhead of benchmark curation~\citep{du2023classeval}, we propose a fully automated construction pipeline (Figure~\ref{fig:Overview}). First, we harvest code snippets across 11 domains from GitHub repositories updated within the past one year~\citep{github2024restapi}. Through AST parsing, candidate classes are structurally merged either within similar domains or across distinct domains. This structural fusion enforces compositional reasoning while synthesizing novel configurations to strictly minimize pretraining exposure~\citep{deshpande2024classlevel}.

Subsequently, an LLM generates a test suite for each skeleton. To ensure rigorous alignment, three LLM-judges~\citep{gu2025surveyllmasajudge} evaluate the skeleton-test compatibility. Only pairs securing a full mark from at least two judges proceed. Finally, a separate LLM synthesizes the solution code. We retain only instances that achieve a 100\% pass rate and $\ge 90\%$ line coverage during execution~\citep{liu2023code}, guaranteeing high-quality evaluation samples~\citep{rahman2025}.

\subsection{Stage 1: Data Collection and Structuring}
ClassEval-Pro expands the original ClassEval taxonomy~\cite{du2023classeval} from 7 to 11 domains, yielding 300 diverse class-level tasks (Table~\ref{Classification}). 

We query the GitHub Search API~\cite{github2024restapi} for Python
repositories created on or after January 1, 2025, ensuring that all collected code postdates current LLM knowledge cutoffs and mitigates data contamination~\cite{jimenez2024swebench}. Keywords drawn from the eleven-domain matrix are issued as disjunctive queries across three star-count tiers (100–500, 500–1,000, $>$1,000), following a stratified sampling strategy akin to The Stack~\cite{kocetkov2022}, with duplicates removed via URL deduplication. Each repository is shallow-cloned, and all non-test Python files are parsed with the \texttt{ast} module; an import-purity analysis discards any file referencing third-party packages or relative imports, retaining only classes whose dependencies resolve exclusively to the standard library—a stricter variant of the self-containedness criterion of CoderEval~\cite{Zhang_2024}. Structural constraints further require at least 5 methods and 40–800 lines of code. From 206 repositories and 1,114 extracted classes, 383 survived all filters.

Surviving candidates undergo three-level verification: (1)~\textit{content integrity}—checking for non-empty content and the absence of HTML/XML artifacts; (2)~\textit{compilation}—parsing with \texttt{ast.parse} and \texttt{compile()} to catch syntax and structural errors; and (3)~\textit{reference completeness}—a scope-aware AST visitor that maintains a scope stack to flag any unresolved name reference, going beyond the parsability filter of StarCoder~\cite{li2023starcoder} to ensure genuine self-containedness. Validated classes are serialized into structured JSON records containing repository metadata, imports, docstrings, and per-method signatures and source code.

To evaluate compositional code generation~\cite{hou2024largelanguagemodelssoftware}, we implement two combination strategies driven by AST-augmented prompts that specify target functionalities and interface constraints~\cite{deshpande2024classlevel}:

\begin{table}[tbp]
\centering
\footnotesize
\caption{Domain Distribution of ClassEval-Pro Tasks}
\label{Classification}
\begin{tabular}{lccc}
\toprule
\textbf{Domain} & \textbf{Same-domain} & \textbf{Cross-domain} & \textbf{Total} \\
\midrule
Finance \& Ecommerce      & 8  & 33 & 41 \\
File Handling             & 8  & 30 & 38 \\
Management Systems        & 18 & 19 & 37 \\
Network \& Web            & 3  & 33 & 36 \\
Database Operations       & 5  & 27 & 32 \\
Security \& Crypto        & 3  & 23 & 26 \\
Natural Lang.\ Process.   & 5  & 18 & 23 \\
Game Development          & 3  & 18 & 21 \\
Mathematical Operations   & 6  & 12 & 18 \\
Utils                     & 6  & 9  & 15 \\
Data Formatting           & 2  & 11 & 13 \\
\midrule
\textbf{Total}            & \textbf{67} & \textbf{233} & \textbf{300} \\
\bottomrule
\end{tabular}
\end{table}

\begin{itemize}[leftmargin=*]
\item \textbf{Intra-Domain Composition:} Merging classes within a single domain. For example, fusing \texttt{SinglyLinkedList} and \texttt{Stack} into a \texttt{LinkedStack} requires the model to optimize internal method calls while maintaining structural integrity.
\item \textbf{Cross-Domain Composition:} Human experts first categorize composable domains based on real-world applications. To simulate complex scenarios, classes are then paired across these domains. For instance, coupling \texttt{SQLQueryBuilder} (Database) with \texttt{Player}\-\texttt{Inventory} (Game) necessitates unified state management and cross-domain functional mapping.
\end{itemize}

As shown in Table~\ref{Comparision}, these compositional strategies yield tasks with significantly higher complexity—quantified by average lines of code (LOC), dependency depth, and method count—exhibiting denser intra-class dependencies than standalone tasks in prior benchmarks~\cite{du2023classeval}.

\begin{table}[t]
\centering
\footnotesize
\caption{Code Metrics Statistics}
\label{Comparision}
\resizebox{\columnwidth}{!}{%
\begin{tabular}{@{}llccccccc@{}}
\toprule
\multirow{2}{*}{\textbf{Dataset}} & \multirow{2}{*}{\textbf{Type}} & \multirow{2}{*}{\textbf{Number}} & \multicolumn{2}{c}{\textbf{LOC}} & \multicolumn{2}{c}{\textbf{Dep.}} & \multicolumn{2}{c}{\textbf{Method}} \\ \cmidrule(lr){4-5} \cmidrule(lr){6-7} \cmidrule(lr){8-9}
 &  &  & Avg & Med & Avg & Med & Avg & Med \\ \midrule
Original & Standard & 100 & 45.7 & 33 & 1.77 & 2.0 & 4.97 & 5.0 \\ \midrule
\multirow{2}{*}{ClassEval-Pro} & Cross-Domain & 233 & 117.0 & 114 & 3.01 & 3.0 & 9.53 & 9.0 \\
 & Same-Domain & 67 & 122.0 & 100 & 2.85 & 3.0 & 8.60 & 8.0 \\ \bottomrule
\end{tabular}%
}
\end{table}

\subsection{Stage 2: Test Generation and Filtering}
After obtaining the combined class skeletons from Stage~1, we provide multi-perspective test case skeletons in the prompt and instruct the LLM to generate matching test code in strict accordance with the class skeleton. The skeleton and generated test code are then submitted to three independent LLM judges ~\cite{gu2025surveyllmasajudge, li2024llmsasjudges}, each of which scores the alignment between the class skeleton and the test code to ensure full compatibility. Only combinations that receive full marks from at least two of the three judges advance to Stage~3. 

\subsection{Stage 3: Solution Code Generation and Excution Verification}
In this stage, the verified class skeletons and their paired test cases are provided to the LLM, which generates complete solution code. The generated solution code and test code then pass through two rounds of filtering. In the first round, an isolated environment is created and compilation is performed; only combinations that compile successfully proceed to the second round. We then use \texttt{pytest} to analyze line coverage of the test cases, retaining only classes whose test coverage reaches at least 90\%. The final dataset comprises class names, class skeletons, test code, corresponding test classes, and solution code. To ensure data integrity, human experts conducted a final verification to confirm that all problem descriptions are accurate and unambiguous.

\section{Experimental Design}

\textbf{Research Questions.} To benchmark LLMs on \textsc{ClassEval-Pro}, we study four complementary research questions covering benchmark difficulty, overall model performance, strategy effects, and failure modes:
\begin{itemize}[leftmargin=*]
    \item \textbf{RQ1:} How effectively does our pipeline increase task difficulty compared to the original ClassEval? This RQ evaluates whether \textsc{ClassEval-Pro} is more challenging in a controlled and meaningful way.
    \item \textbf{RQ2:} How do state-of-the-art LLMs perform on \textsc{ClassEval-Pro} under holistic generation? This RQ provides the main benchmark comparison under a unified setting.
    \item \textbf{RQ3:} How do different generation strategies affect class-level code generation performance across models? This RQ examines the sensitivity of \textsc{ClassEval-Pro} to different class construction strategies.
    \item \textbf{RQ4:} What types of errors dominate LLM failures on class-level code generation? This RQ identifies the main failure patterns exposed by \textsc{ClassEval-Pro}.
\end{itemize}

\subsection{Studied LLMs}

We study five LLMs, covering both proprietary and open-weight systems. Specifically, our evaluation includes two commercial frontier models, GPT-5.1 and Gemini-2.5 Pro, together with three open-weight instruction-tuned MoE models, Kimi-K2, Qwen3-480B-A35B, and Qwen3-30B-A3B. These models span a wide range of scales, from 30.5B parameters to trillion-scale MoE architectures, with active parameter sizes ranging from 3.3B to 35B. For the open-weight models, the officially released specifications indicate training corpora of 15.5--36 trillion tokens and context windows ranging from 32k to 256k tokens, while the proprietary models provide comparably strong long-context support but do not publicly disclose parameter size or training-token details. This selection enables a balanced comparison between closed and open-weight models, while also supporting within-family scaling analysis through the two Qwen3 variants.

\subsection{Studied Generation Strategies}
For class-level code generation~\cite{du2023classeval}, we investigate five generation strategies that span the spectrum from single-pass synthesis to structured decomposition and iterative refinement~\cite{shi2024code}.

\begin{itemize}[leftmargin=*]
\item {Holistic}: Given a code class skeleton as input, the LLM is required to generate the complete class implementation in a single pass~\citep{du2023classeval}. 

\item {Incremental}: Following~\citet{du2023classeval}, classes are generated by iteratively predicting methods in their declaration order. Each step leverages the previously generated code as context, continuing until class completion.

\item {Compositional}: Import statements, class definitions, and method bodies are generated independently from the skeleton~\citep{du2023classeval}. Each fragment is produced without inter-method context, then assembled to form the final class.

\item {Bottom-Up}: Utilizing a two-stage dependency-guided framework~\citep{chen2024premise}, this approach categorizes methods into hierarchical levels based on call graphs and shared states. Methods are generated incrementally from the lowest dependency level to the highest.

\item {Top-Down}: This approach employs the same dependency analysis as Bottom-Up but inverts the implementation sequence, prioritizing high-level methods before their lower-level dependencies.

\end{itemize} 

We provide one example for each generation strategy in the released code repository for clearer illustration.

\subsection{Evaluation Metrics}

To evaluate generation correctness, we follow established literature by adopting  \text{Pass@}k as our primary metric.


To ensure a reliable evaluation, the underlying benchmark is rigorously validated during the dataset construction phase. We automatically execute candidate solutions against the test suites in an isolated environment. A generation task is included in the final dataset only if its reference solution satisfies strict execution criteria: (1) passing all method-level and class-level tests; (2) achieving a statement coverage strictly greater than 90\%; and (3) completing execution within a 60-second timeout limit to prevent unbounded resource consumption.

To maintain practical computational overhead and response times, we set n to five. To mitigate high sampling variance, we employ the unbiased estimator consistent with prior work~\cite{du2023classeval}.



\subsection{Implementation Details}
We evaluate our benchmark using five models: GPT-5.1, Gemini-2.5-Pro, Qwen3-480B-A35B, Qwen3-30B-A3B and Kimi-K2. All models are accessed via their respective APIs. For all generation tasks, we set temperature=0.2 and max\_tokens=16438. To compute Pass@$k$, we sample $k=5$ candidate solutions per task. Our automated evaluation pipeline, including code execution and test verification via pytest, runs on a local server equipped with 8$\times$ NVIDIA A800 (80GB) GPUs. We enforce a strict 60-second timeout per task to bound resource consumption. Due to API inference cost constraints, all models are evaluated with a single run. To ensure reproducibility, all constructed data, generated outputs, and evaluation scripts will be open-sourced upon acceptance.

\section{Experimental Results}

\begin{table*}[t]
 \caption{Results of different generation strategies with various models}
 \label{tab:hierarchical-strategies}
 \centering
 \resizebox{\textwidth}{!}{%
 \begin{tabular}{l|l|ccc|ccc|ccc|ccc}
 \toprule
 \multirow{2}{*}{Model} & \multirow{2}{*}{Generation Strategy} & \multicolumn{6}{c|}{Class-level} & \multicolumn{6}{c}{Method-level} \\
 \cmidrule(lr){3-8} \cmidrule(lr){9-14}
 & & \multicolumn{3}{c|}{Full Success} & \multicolumn{3}{c|}{Partial Success} & \multicolumn{3}{c|}{Full Success} & \multicolumn{3}{c}{Partial Success} \\
 \cmidrule(lr){3-5} \cmidrule(lr){6-8} \cmidrule(lr){9-11} \cmidrule(lr){12-14}
 & & Pass@1 & Pass@3 & Pass@5 & Pass@1 & Pass@3 & Pass@5 & Pass@1 & Pass@3 & Pass@5 & Pass@1 & Pass@3 & Pass@5 \\
 \midrule
 \multirow{5}{*}{\textsc{GPT-5.1}} & Holistic & $27.9\%$ & $35.8\%$ & $39.0\%$ & $47.5\%$ & $55.1\%$ & $57.3\%$ & $61.3\%$ & $67.3\%$ & $69.0\%$ & $70.0\%$ & $74.9\%$ & $76.1\%$ \\
 & Incremental & $20.9\%$ & $26.0\%$ & $28.0\%$ & $33.3\%$ & $39.5\%$ & $41.7\%$ & $43.6\%$ & $49.0\%$ & $50.8\%$ & $51.0\%$ & $55.4\%$ & $56.8\%$ \\
 & Compositional & $17.7\%$ & $23.4\%$ & $26.3\%$ & $30.5\%$ & $38.5\%$ & $42.3\%$ & $40.1\%$ & $46.3\%$ & $48.8\%$ & $48.1\%$ & $54.0\%$ & $56.3\%$ \\
 & Top-down & $34.7\%$ & $47.9\%$ & $53.0\%$ & $53.1\%$ & $66.2\%$ & $70.0\%$ & $66.8\%$ & $79.7\%$ & $83.1\%$ & $73.8\%$ & $85.4\%$ & $88.1\%$ \\
 & Bottom-up & $37.3\%$ & $47.2\%$ & $51.3\%$ & $56.9\%$ & $64.7\%$ & $68.3\%$ & $73.0\%$ & $78.7\%$ & $80.8\%$ & $80.4\%$ & $84.4\%$ & $85.9\%$ \\
 \midrule
 \multirow{5}{*}{\textsc{Gemini-2.5-Pro}} & Holistic & $44.7\%$ & $58.1\%$ & $62.7\%$ & $69.3\%$ & $78.6\%$ & $80.7\%$ & $80.5\%$ & $88.1\%$ & $90.3\%$ & $89.1\%$ & $93.7\%$ & $94.8\%$ \\
 & Incremental & $45.4\%$ & $57.4\%$ & $61.7\%$ & $66.5\%$ & $76.5\%$ & $78.3\%$ & $77.1\%$ & $86.0\%$ & $87.9\%$ & $84.4\%$ & $91.1\%$ & $92.2\%$ \\
 & Compositional & $1.3\%$ & $2.6\%$ & $3.0\%$ & $1.8\%$ & $3.3\%$ & $3.7\%$ & $8.0\%$ & $15.4\%$ & $19.4\%$ & $19.2\%$ & $32.2\%$ & $39.0\%$ \\
 & Top-down & $48.4\%$ & $63.0\%$ & $68.0\%$ & $69.9\%$ & $80.9\%$ & $84.7\%$ & $82.2\%$ & $90.4\%$ & $92.2\%$ & $89.5\%$ & $95.0\%$ & $96.0\%$ \\
& Bottom-up & $45.5\%$ & $59.5\%$ & $64.0\%$ & $66.7\%$ & $79.0\%$ & $81.7\%$ & $76.7\%$ & $88.1\%$ & $90.9\%$ & $83.4\%$ & $93.0\%$ & $94.8\%$ \\
 \midrule
 \multirow{5}{*}{\textsc{qwen3-480b-a35b}} & Holistic & $45.6\%$ & $52.9\%$ & $56.3\%$ & $74.1\%$ & $79.4\%$ & $81.0\%$ & $84.2\%$ & $87.5\%$ & $88.7\%$ & $92.6\%$ & $94.4\%$ & $95.0\%$ \\
 & Incremental & $37.9\%$ & $50.9\%$ & $56.0\%$ & $54.9\%$ & $70.0\%$ & $76.0\%$ & $63.6\%$ & $77.8\%$ & $83.0\%$ & $69.0\%$ & $83.1\%$ & $88.3\%$ \\
 & Compositional & $13.5\%$ & $25.2\%$ & $30.3\%$ & $31.1\%$ & $52.7\%$ & $61.0\%$ & $48.9\%$ & $68.8\%$ & $74.3\%$ & $65.5\%$ & $83.4\%$ & $86.9\%$ \\
 & Top-down & $45.3\%$ & $55.1\%$ & $58.7\%$ & $73.9\%$ & $81.1\%$ & $83.3\%$ & $83.5\%$ & $88.7\%$ & $90.2\%$ & $91.6\%$ & $94.8\%$ & $95.5\%$ \\
 & Bottom-up & $46.3\%$ & $54.1\%$ & $56.3\%$ & $72.7\%$ & $77.0\%$ & $78.3\%$ & $82.7\%$ & $86.2\%$ & $87.4\%$ & $90.2\%$ & $92.2\%$ & $93.0\%$ \\
  \midrule
 \multirow{5}{*}{\textsc{qwen3-30b-a3b}} & Holistic & $40.5\%$ & $51.1\%$ & $55.7\%$ & $67.3\%$ & $77.2\%$ & $81.0\%$ & $78.2\%$ & $85.6\%$ & $88.0\%$ & $87.9\%$ & $92.9\%$ & $94.3\%$ \\
 & Incremental & $38.3\%$ & $47.1\%$ & $50.7\%$ & $64.3\%$ & $73.2\%$ & $76.0\%$ & $76.2\%$ & $83.7\%$ & $85.5\%$ & $85.7\%$ & $91.4\%$ & $92.5\%$ \\
 & Compositional & $7.5\%$ & $17.1\%$ & $22.7\%$ & $22.8\%$ & $42.2\%$ & $50.3\%$ & $38.8\%$ & $60.6\%$ & $67.2\%$ & $55.6\%$ & $76.3\%$ & $81.4\%$ \\
 & Top-down & $35.3\%$ & $49.0\%$ & $55.0\%$ & $61.9\%$ & $74.4\%$ & $78.0\%$ & $73.8\%$ & $84.0\%$ & $87.0\%$ & $83.4\%$ & $91.2\%$ & $93.1\%$ \\
 & Bottom-up & $38.6\%$ & $51.0\%$ & $55.3\%$ & $67.5\%$ & $76.6\%$ & $79.0\%$ & $77.1\%$ & $85.3\%$ & $87.4\%$ & $86.8\%$ & $92.6\%$ & $93.8\%$\\
 \midrule
\multirow{5}{*}{\textsc{Kimi-K2}} & Holistic & $45.1\%$ & $53.1\%$ & $56.0\%$ & $72.1\%$ & $78.5\%$ & $81.0\%$ & $83.0\%$ & $87.3\%$ & $88.7\%$ & $91.0\%$ & $93.5\%$ & $94.3\%$ \\
 & Incremental & $32.7\%$ & $45.2\%$ & $49.3\%$ & $52.3\%$ & $65.3\%$ & $68.7\%$ & $61.9\%$ & $73.4\%$ & $76.2\%$ & $68.6\%$ & $78.8\%$ & $81.0\%$ \\
 & Compositional & $21.7\%$ & $31.6\%$ & $34.0\%$ & $41.5\%$ & $54.0\%$ & $57.0\%$ & $56.1\%$ & $67.5\%$ & $70.2\%$ & $69.6\%$ & $79.0\%$ & $81.0\%$ \\
 & Top-down & $44.5\%$ & $52.5\%$ & $55.0\%$ & $70.8\%$ & $78.0\%$ & $80.3\%$ & $82.9\%$ & $87.5\%$ & $88.7\%$ & $91.1\%$ & $93.7\%$ & $94.4\%$ \\
 & Bottom-up & $31.5\%$ & $46.7\%$ & $54.0\%$ & $49.7\%$ & $68.7\%$ & $76.0\%$ & $59.1\%$ & $78.6\%$ & $85.7\%$ & $65.0\%$ & $84.7\%$ & $91.6\%$ \\
 \bottomrule
 \end{tabular}
 }
\end{table*}

\subsection{RQ1: Dataset Difficulty}

To address RQ1, we evaluate task difficulty across two dimensions: structural complexity and semantic diversity.

\begin{figure}[t]
    \includegraphics[width=0.85\linewidth]{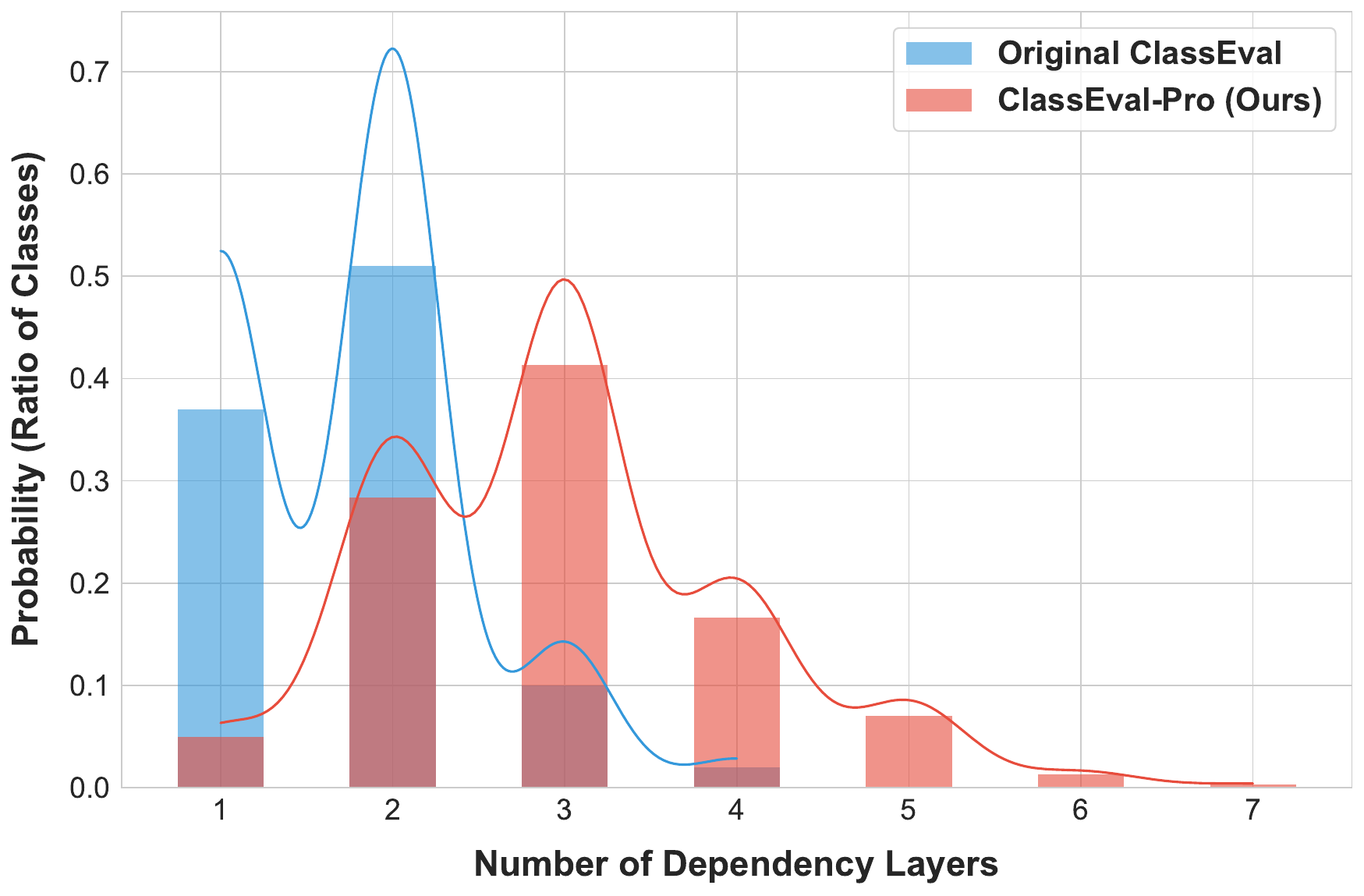}
    \caption{Normalized Distribution of Dependency Layers}
    \label{fig:dep_layers}

    \vspace{1em}

    \includegraphics[width=0.85\linewidth]{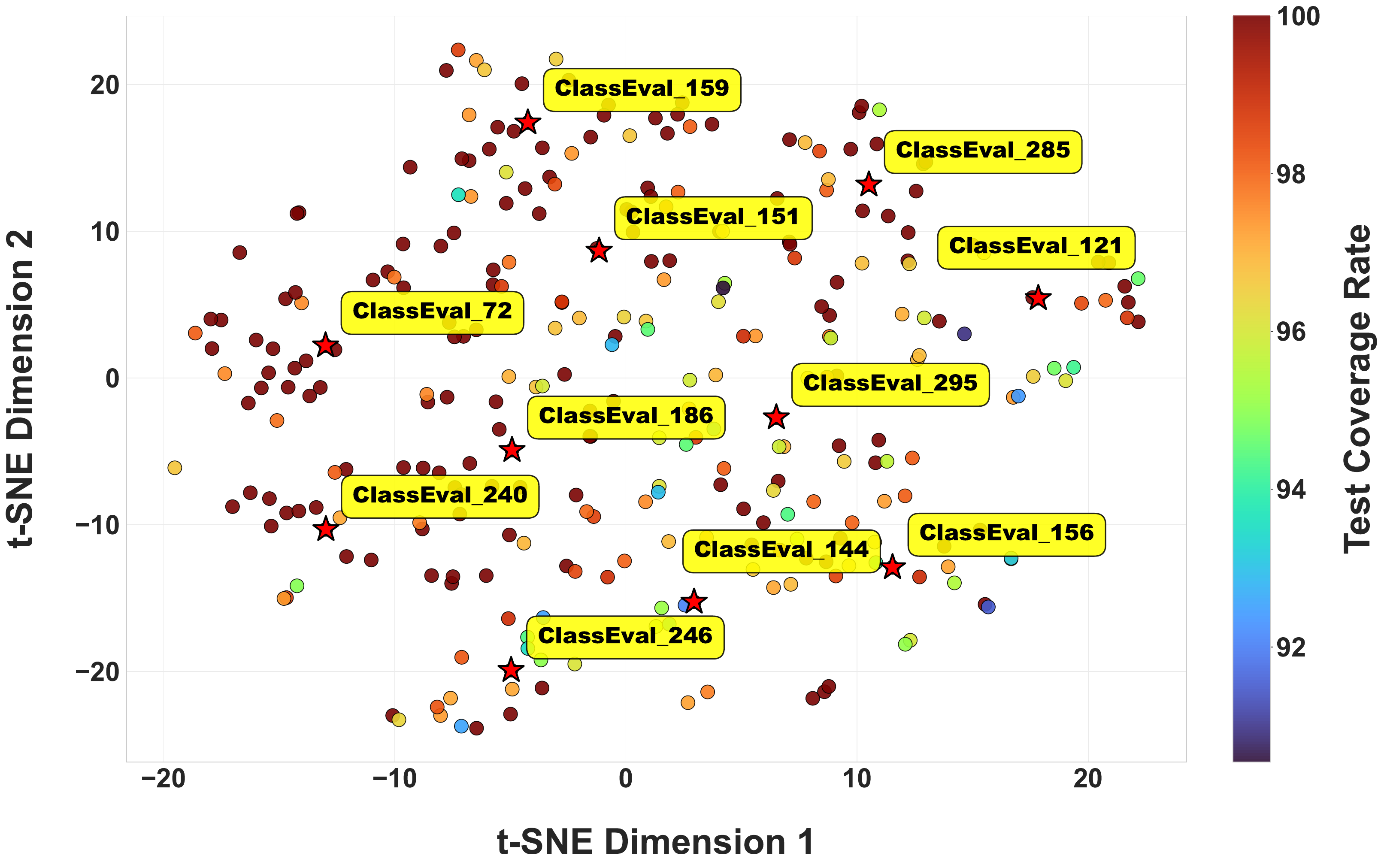}
    \caption{Semantic Landscape of ClassEval-Pro}
    \label{fig:tsne}
    \vspace{-0.2cm}
\end{figure}

\textbf{Structural Complexity:} Using internal method dependencies as a proxy (Figure~\ref{fig:dep_layers}), \textsc{ClassEval-Pro} exhibits a significant rightward shift. Unlike the original ClassEval, which peaks at 1--2 layers, our benchmark centers at 3 layers with a long tail extending to 7. This deeper hierarchy amplifies difficulty by strictly requiring multi-step compositional reasoning rather than localized function completion.

\textbf{Semantic Diversity and Rigor:} The t-SNE projection of CodeBERT embeddings (Figure~\ref{fig:tsne}) displays wide dispersion, validating that our combination strategies yield a highly diverse task space.

\vspace{-0.3em}
\begin{boxK}
\small \faIcon{pencil-alt} \textbf{Finding 1:} Our automated pipeline increases benchmark difficulty along both structural and semantic dimensions without sacrificing evaluation rigor. As a result, ClassEval-Pro provides a more challenging and more discriminative testbed for class-level code generation than the original ClassEval.
\end{boxK}
\vspace{-0.3em}

\subsection{RQ2: Overall Performance Analysis}

We first evaluate all models under \textit{holistic} generation, where each model generates the entire target class in a single pass. This setting serves as a unified baseline for comparing models because it avoids introducing additional decomposition, ordering, or repair mechanisms, and therefore directly reflects each model's intrinsic ability to plan and synthesize a complete class-level artifact from the given specification and skeleton.

\textbf{Comparison among LLMs.} Under holistic generation, clear capability differences emerge across models. \textsc{Qwen3-480B-A35B} achieves the strongest Pass@1 Class-level Full Success ($45.6\%$), followed closely by \textsc{Kimi-K2} ($45.1\%$) and \textsc{Gemini-2.5-Pro} ($44.7\%$). In contrast, \textsc{GPT-5.1} lags substantially behind at $27.9\%$. The resulting $17.7$-point gap between the best and worst models confirms that ClassEval-Pro has sufficient discriminative power even among recent frontier systems. The gap is not limited to exact class-level correctness: the same ranking trend is broadly reflected in Class-level Partial Success and Method-level metrics, suggesting that stronger models are not only more likely to complete the entire class correctly, but also produce more locally correct method implementations.

The distinction between class-level and method-level correctness further reveals the compositional nature of the benchmark. For example, under holistic generation, \textsc{Qwen3-480B-A35B} reaches $84.2\%$ Method-level Full Success at Pass@1, but only $45.6\%$ Class-level Full Success. Similarly, \textsc{Kimi-K2} achieves $83.0\%$ Method-level Full Success but $45.1\%$ Class-level Full Success, while \textsc{Gemini-2.5-Pro} achieves $80.5\%$ versus $44.7\%$. This consistent gap indicates that many generated solutions contain individually correct methods but still fail to satisfy the complete class-level contract. In other words, ClassEval-Pro exposes failures in cross-method coordination, state consistency, and interface preservation that would be obscured if evaluation only considered method-level correctness.

\textbf{Effect of model scale and sampling budget.} A clear scaling effect is evident within the Qwen family. The larger \textsc{Qwen3-480B-A35B} outperforms \textsc{Qwen3-30B-A3B} by $5.1$ points on Pass@1 Class-level Full Success ($45.6\%$ vs.\ $40.5\%$), and the advantage also appears in Method-level Full Success ($84.2\%$ vs.\ $78.2\%$) and Method-level Partial Success ($92.6\%$ vs.\ $87.9\%$). These results suggest that larger model capacity remains beneficial for class-level generation, especially when the task requires maintaining long-range dependencies across multiple methods.

Increasing the sampling budget improves all models, but the magnitude of improvement varies substantially. \textsc{Gemini-2.5-Pro} shows the largest Pass@1-to-Pass@5 gain in Class-level Full Success, rising from $44.7\%$ to $62.7\%$ ($+18.0$ points). \textsc{GPT-5.1} also improves from $27.9\%$ to $39.0\%$ ($+11.1$ points), but remains far below the leading models even at Pass@5. By contrast, \textsc{Qwen3-480B-A35B} and \textsc{Kimi-K2} show more moderate gains, increasing by $10.7$ and $10.9$ points, respectively. This suggests that some models, especially \textsc{Gemini-2.5-Pro}, benefit more from sampling diversity, whereas others produce more stable but less variable candidates under the same holistic prompt.

\textbf{Partial success and remaining bottlenecks.} Partial Success scores are consistently higher than Full Success scores at both class and method levels. For instance, \textsc{Qwen3-480B-A35B} achieves $74.1\%$ Class-level Partial Success at Pass@1, compared with $45.6\%$ Class-level Full Success; \textsc{Kimi-K2} similarly reaches $72.1\%$ versus $45.1\%$. This gap shows that many generated classes are not entirely incorrect, but fail on a subset of methods, edge cases, or interaction paths. The highest Method-level Partial Success scores are also above $90\%$ for the strongest models, indicating that local method synthesis is often largely successful. The remaining performance bottleneck therefore lies less in producing syntactically valid or locally plausible code, and more in composing these pieces into a coherent class that satisfies all tests simultaneously.

Overall, holistic generation reveals three key observations. First, ClassEval-Pro clearly distinguishes model capabilities, with top-performing models clustering around $45\%$ Pass@1 Class-level Full Success and weaker models falling far behind. Second, larger models retain an advantage within the same model family, supporting the importance of capacity for long-range dependency planning. Third, the large gap between method-level and class-level success demonstrates that class-level code generation remains a challenging compositional problem: models often implement individual methods correctly, but fail to maintain global consistency across the generated class.

\vspace{-0.3em}
\begin{boxK}
\small \faIcon{pencil-alt} \textbf{Finding 2:} Under holistic generation, ClassEval-Pro clearly differentiates frontier LLMs: the strongest models achieve around $45\%$ Pass@1 Class-level Full Success, while weaker models lag substantially behind. The large gap between method-level and class-level success further shows that the main bottleneck is no longer isolated method implementation, but maintaining cross-method dependencies, shared state, and interface consistency across the full class.
\end{boxK}
\vspace{-0.3em}

\subsection{RQ3: Impact of Generation Strategies}

Different generation strategies may favor different combinations of model capability and task complexity. We therefore evaluate five representative strategies---\textit{holistic}, \textit{incremental}, \textit{compositional}, \textit{top-down}, and \textit{bottom-up}---to examine how strategy choice affects class-level code generation performance.

As shown in Table~\ref{tab:hierarchical-strategies}, strategy effectiveness is strongly model-dependent, and no single strategy dominates across all settings. Nevertheless, several clear trends emerge. First, structured dependency-aware strategies are often beneficial when they help the model preserve inter-method relationships. For \textsc{GPT-5.1}, \textit{bottom-up} achieves the best Pass@1 Class-level Full Success ($37.3\%$), outperforming both \textit{holistic} ($27.9\%$) and \textit{top-down} ($34.7\%$). The improvement is even more visible at Method-level Full Success, where \textit{bottom-up} reaches $73.0\%$ compared with $61.3\%$ under \textit{holistic}. This suggests that implementing lower-level or dependency-light methods first can help weaker holistic planners build a more stable foundation for subsequent class construction.

Second, \textit{top-down} is particularly effective for models that already possess strong global planning ability. On \textsc{Gemini-2.5-Pro}, \textit{top-down} achieves the best performance across most class-level metrics, improving Pass@1 Class-level Full Success from $44.7\%$ under \textit{holistic} to $48.4\%$, and Pass@5 from $62.7\%$ to $68.0\%$. It also improves Class-level Partial Success at Pass@5 from $80.7\%$ to $84.7\%$. This indicates that explicitly planning high-level class behavior before filling in implementation details can help strong models better preserve whole-class intent and interface consistency.

However, the benefit of structured generation is not universal. For \textsc{Qwen3-480B-A35B}, \textit{holistic}, \textit{top-down}, and \textit{bottom-up} perform very closely: Pass@1 Class-level Full Success is $45.6\%$, $45.3\%$, and $46.3\%$, respectively. This suggests that very strong open-weight models can already internalize much of the dependency structure in a single-pass setting, leaving limited room for explicit decomposition to improve exact class-level correctness. For \textsc{Qwen3-30B-A3B}, \textit{holistic} remains the best Pass@1 strategy ($40.5\%$), while \textit{bottom-up} ($38.6\%$), \textit{incremental} ($38.3\%$), and \textit{top-down} ($35.3\%$) do not provide consistent gains. Similarly, on \textsc{Kimi-K2}, \textit{holistic} ($45.1\%$) and \textit{top-down} ($44.5\%$) remain strongest, whereas \textit{bottom-up} drops substantially to $31.5\%$. These cases show that decomposition can also introduce additional failure points when intermediate dependency analysis or method ordering does not align with the model's own generation dynamics.

The weakest strategy overall is \textit{compositional} generation. Because it generates different class components in isolation and assembles them afterwards, it often fails to preserve shared state, method contracts, and cross-method assumptions. This is especially severe for \textsc{Gemini-2.5-Pro}, where Pass@1 Class-level Full Success collapses to only $1.3\%$, compared with $44.7\%$ under \textit{holistic}. Similar degradation appears on \textsc{Qwen3-30B-A3B} ($7.5\%$), \textsc{Qwen3-480B-A35B} ($13.5\%$), and \textsc{GPT-5.1} ($17.7\%$). Although \textit{compositional} sometimes retains moderate Method-level Partial Success, its class-level results confirm that locally plausible fragments are insufficient when the final artifact must satisfy global class consistency.

\textit{Incremental} generation is more stable than \textit{compositional}, but still generally weaker than \textit{holistic} or dependency-aware structured strategies. For example, it reduces \textsc{GPT-5.1}'s Pass@1 Class-level Full Success from $27.9\%$ to $20.9\%$, \textsc{Qwen3-480B-A35B}'s from $45.6\%$ to $37.9\%$, and \textsc{Kimi-K2}'s from $45.1\%$ to $32.7\%$. The likely reason is that method-by-method generation can preserve some local context, but it gradually accumulates inconsistencies as earlier generated methods constrain later ones. The only clear exception is \textsc{Gemini-2.5-Pro}, where \textit{incremental} achieves a Pass@1 Class-level Full Success of $45.4\%$, slightly above \textit{holistic} ($44.7\%$), although \textit{top-down} still performs best.

Overall, strategy choice substantially changes class-level generation outcomes. Structured decomposition can improve performance when it provides useful dependency guidance, as seen with \textsc{GPT-5.1} and \textsc{Gemini-2.5-Pro}. Yet decomposition is not automatically beneficial: overly fragmented generation can destroy whole-class coherence, and even dependency-aware ordering may hurt models whose single-pass generation already captures the class structure well. These results show that ClassEval-Pro is sensitive not only to model capability, but also to how the generation process organizes dependencies, shared state, and interface contracts.

\vspace{-0.3em}
\begin{boxK}
\small \faIcon{pencil-alt} \textbf{Finding 3:} Strategy effectiveness depends strongly on model capability and decomposition style. Dependency-aware strategies such as top-down and bottom-up can improve class-level correctness, but excessive or poorly aligned decomposition often harms whole-class coherence. This makes ClassEval-Pro effective at exposing failures in dependency coordination, shared-state consistency, and class construction order.
\end{boxK}
\vspace{-0.3em}

\begin{figure}[t]
    \centering
    \includegraphics[width=0.9\linewidth]{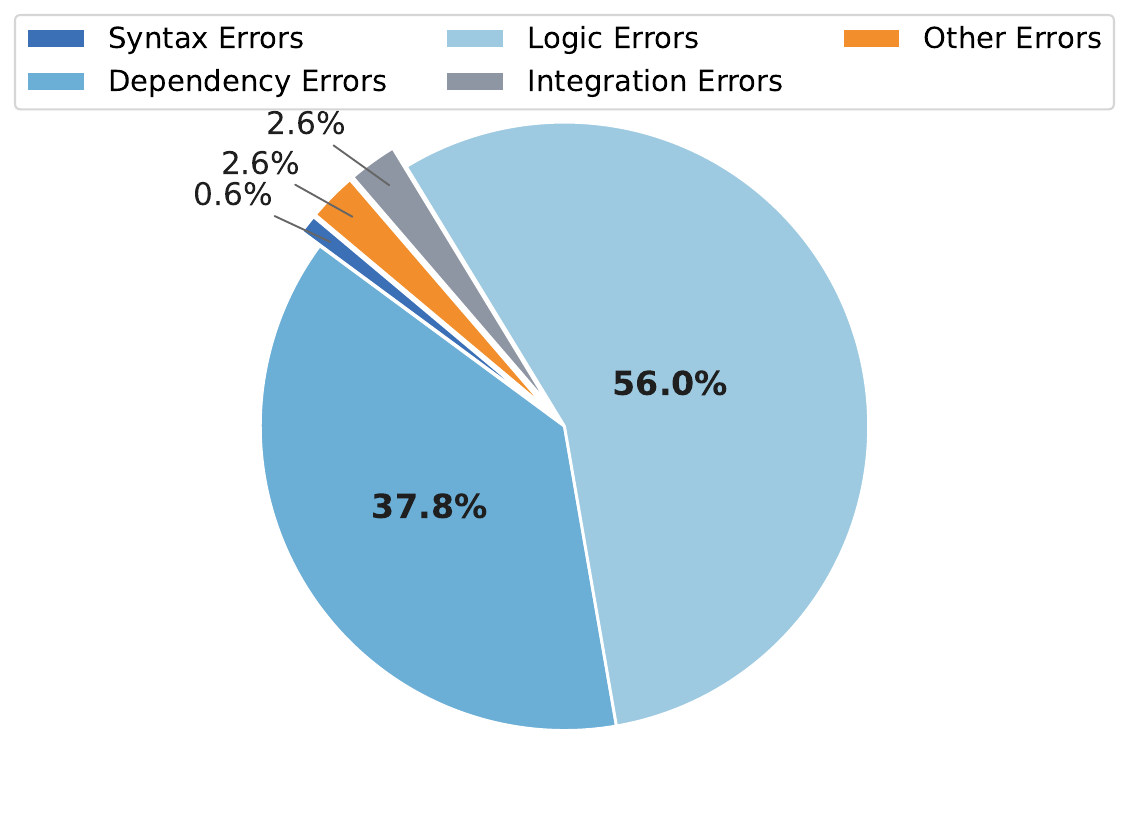}
    \caption{Overall distribution of primary error categories}
    \label{fig:error_distribution_overall}
\end{figure}

\subsection{RQ4: Error Analysis}

To understand the main remaining bottlenecks in class-level code generation, we analyze failed cases across five generation strategies and five frontier LLMs, using them to identify recurring error patterns at both the model and strategy levels.
\subsubsection{Methodology}
We focus on partially failed cases, namely those with non-zero method-level success but class-level failure, as they provide the clearest diagnostic signal. For each strategy, we select the best-performing model as the representative source and apply stratified random sampling across the five test categories, yielding 100 cases per strategy and 500 manually annotated cases in total.

The analysis proceeds in two stages. We first use automated log extraction to collect tracebacks, exception types, and failed test names for coarse categorization. We then conduct manual thematic annotation on the 500-case sample with three annotators who each have more than three years of Python experience, examining prompts, generated code, and tracebacks to identify root causes, with multiple labels allowed when defects co-occur. The taxonomy is developed iteratively on a pilot subset and refined before the main round. Inter-annotator agreement, measured by Fleiss' $\kappa$ over top-level categories, reaches substantial agreement, and all remaining disagreements are resolved through consensus review.

\subsubsection{Error Taxonomy}

We organize consensus annotations into five primary error categories and one benchmark-specific cross-cutting tag. Annotators may assign multiple labels during review, but for aggregate reporting each failed case is mapped to its most direct root cause, with the cross-cutting tag retained separately when needed. 

Concretely, \textbf{Syntax Errors} capture low-level code invalidity such as unresolved names or malformed symbol references; \textbf{Dependency Errors} arise when the generated class fails to preserve required field, method, or library dependencies; \textbf{Logic Errors} describe runnable but semantically incorrect implementations; \textbf{Integration Errors} reflect class-level inconsistencies where individually plausible methods do not function coherently together; and \textbf{Other Errors} cover remaining runnable failures, mainly unsupported or hallucinated behavior beyond the specification.


\subsubsection{Error Distribution Analysis}

Figure~\ref{fig:error_distribution_overall} shows that the dominant bottleneck among failed-but-runnable cases lies in semantic correctness rather than basic executability. For readability, slices below 1\% are slightly enlarged in the figure, although the reported percentages remain unchanged. Across the 500 annotated cases, \textbf{Logic Errors} constitute the largest share at \textbf{56.2\%}, followed by \textbf{Dependency Errors} at \textbf{38.0\%}, whereas \textbf{Syntax Errors} account for only \textbf{0.6\%}. \textbf{Integration Errors} and \textbf{Other Errors} each remain limited at \textbf{2.6\%}. Taken together, these results indicate that once modern LLMs produce runnable class code, their main difficulty is no longer surface-level syntax, but preserving correct semantics, inter-method consistency, and dependency structure. 

Overall, the error distribution suggests that class-level code generation is primarily constrained by reasoning and coordination failures rather than by low-level code formation. In particular, the clear dominance of logic and dependency errors implies that improving benchmark performance will depend less on further reducing syntax mistakes and more on strengthening models' ability to maintain coherent class semantics, accurate method interactions, and reliable use of required state or libraries.


\subsubsection{Error Pattern Analysis by Strategy}

\begin{figure}[t]
    \centering
    \includegraphics[width=0.95\linewidth]{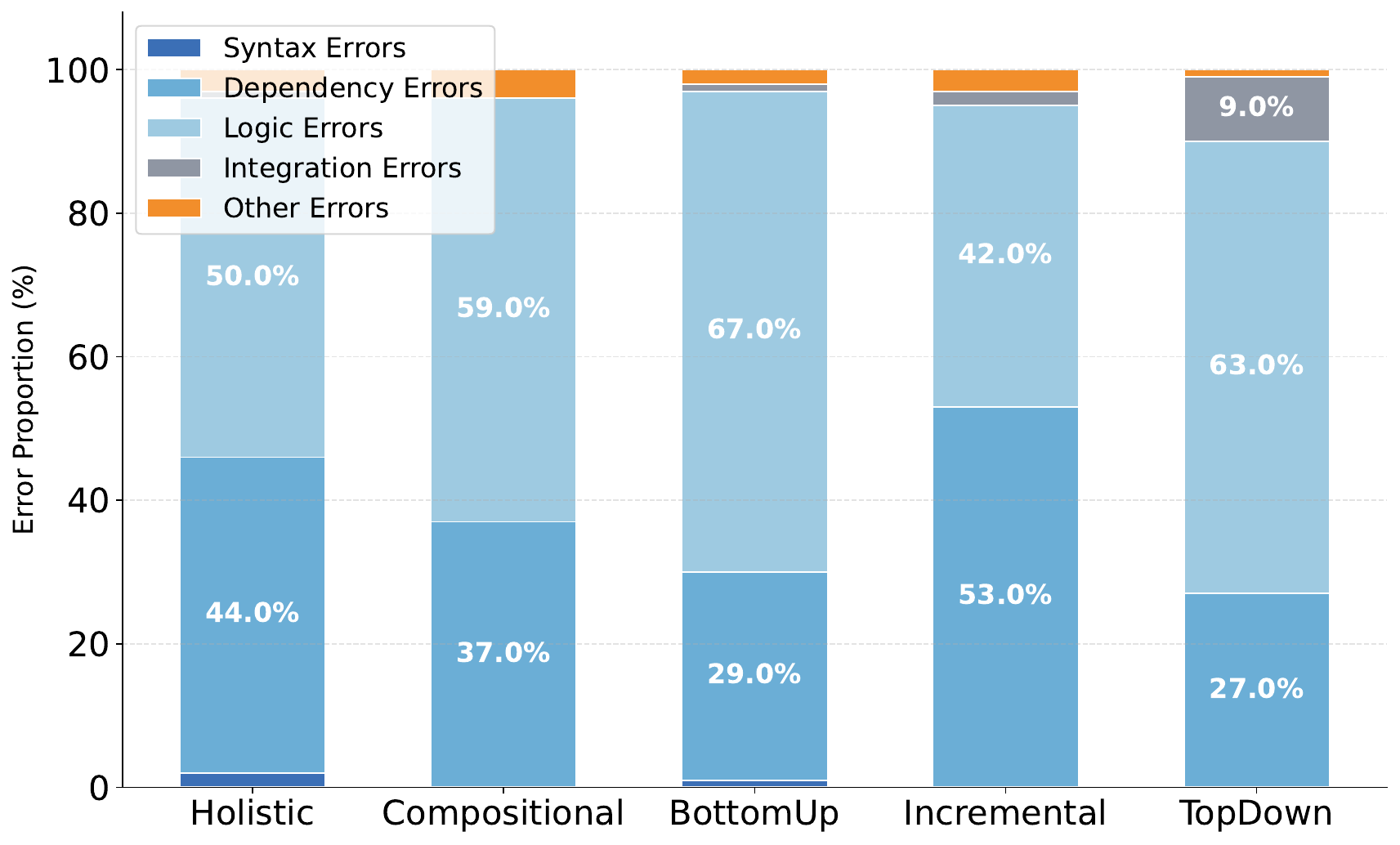}
    \caption{Error distribution across generation strategies}
    \vspace{-0.1cm}
    \label{fig:error_pattern_by_strategy}
\end{figure}

Figure~\ref{fig:error_pattern_by_strategy} shows that strategies fail in systematically different ways. \textsc{Holistic} has a balanced profile: \textbf{Logic Errors} (\textbf{50.0\%}) and \textbf{Dependency Errors} (\textbf{44.0\%}) occur at comparable rates, suggesting full-class generation retains global context but struggles with both semantic correctness and dependency consistency. \textsc{Incremental} has the highest \textbf{Dependency Errors} (\textbf{53.0\%}), indicating step-wise generation is especially vulnerable to broken signatures, missing fields, and cross-method context fragmentation. \textsc{BottomUp} and \textsc{TopDown} shift failures toward \textbf{Logic Errors} (\textbf{67.0\%} and \textbf{63.0\%}, respectively) while keeping dependency errors lower, suggesting hierarchical structuring preserves interfaces but does not resolve semantic reasoning at the class level. \textsc{TopDown} also has the largest \textbf{Integration Errors} share (\textbf{9.0\%}), implying high-level planning can leave inconsistencies in state evolution and cross-method coordination. \textsc{Compositional} shows a mixed profile with \textbf{Logic Errors} at \textbf{59.0\%} and \textbf{Dependency Errors} at \textbf{37.0\%}, indicating independently generated fragments remain hard to reconcile into a coherent class.

Overall, these results reveal a trade-off between dependency preservation and semantic correctness. Strategies with explicit decomposition (\textsc{BottomUp}, \textsc{TopDown}) alleviate dependency failures but shift the burden toward logic and integration problems, whereas \textsc{Incremental} suffers most from context fragmentation. Generation strategy thus changes \emph{how} models fail rather than eliminating failure: unstructured generation exposes dependency drift, while structured generation reveals unresolved semantic composition and class-level coordination challenges.

\subsubsection{Concrete Error Examples}

\begin{figure*}[t]
    \centering
    \includegraphics[width=0.8\textwidth]{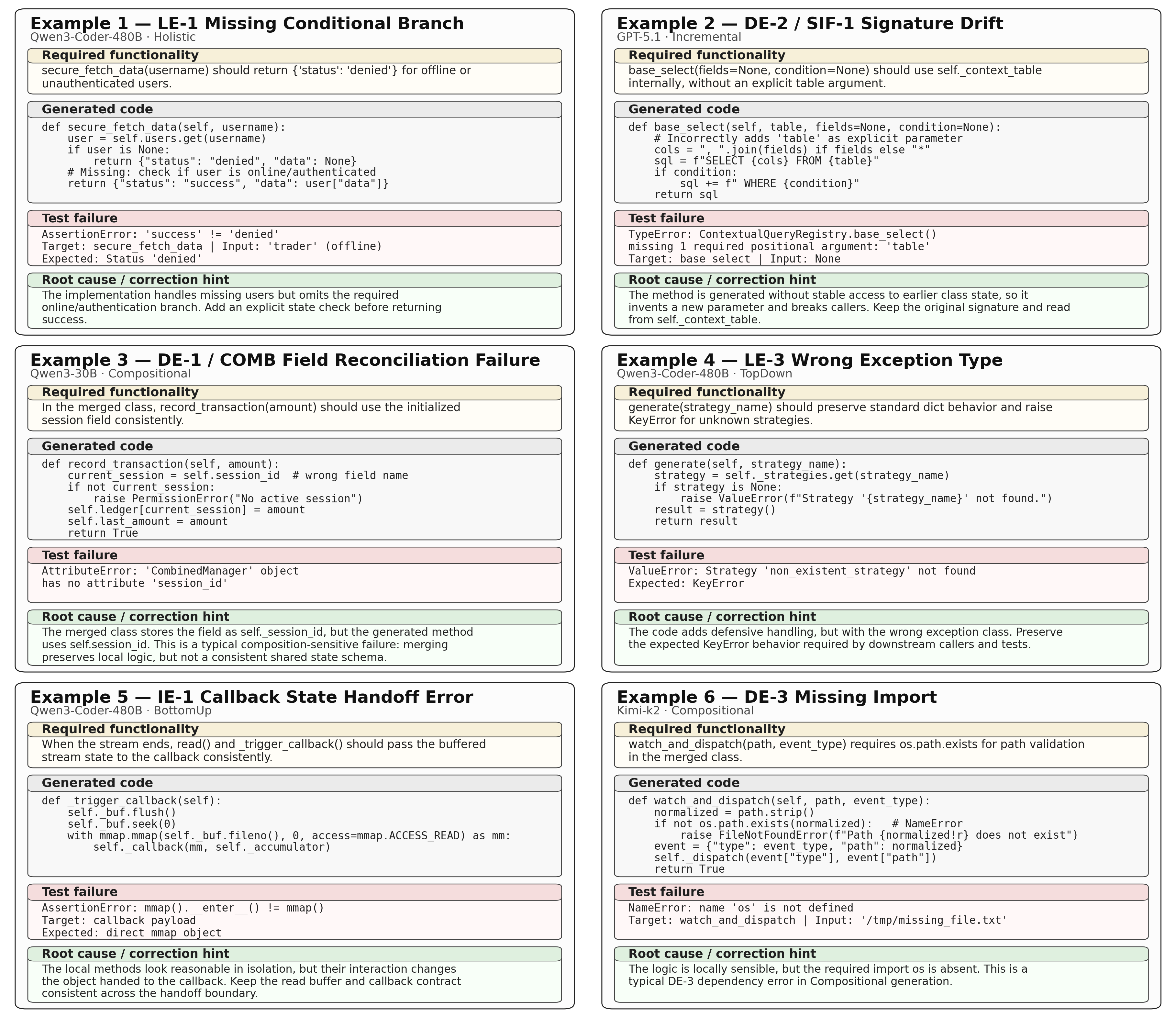}
    \caption{Representative concrete error examples across major error categories. Each example shows the required functionality, generated code, observed test failure, and the corresponding root cause or correction hint.}
    \label{fig:concrete_error_examples}
\end{figure*}

Figure~\ref{fig:concrete_error_examples} presents six representative failure cases covering the major error patterns identified in our taxonomy. Each example includes the required functionality, the generated implementation, the resulting test failure, and a concise root-cause explanation. These examples illustrate that failures in ClassEval-Pro are often not caused by uncompilable or obviously invalid code. Instead, the generated methods are frequently locally plausible but fail to satisfy the broader class-level contract, especially when the implementation must preserve hidden state, method interfaces, exception behaviour, or cross-method data flow.

Example~1 shows a \textbf{Logic Error} caused by a missing conditional branch. The model correctly handles the case where the fetched user is absent, but omits the required branch for offline or unauthenticated users. As a result, an offline user is incorrectly treated as successful rather than denied. This case reflects a common semantic failure mode: the implementation covers the most obvious path but fails to model an important edge case specified by the task.

Example~2 illustrates a \textbf{Dependency Error} with a strategy-induced interface failure. The required method \texttt{base\_select} takes only \texttt{fields} and 
\texttt{condition} as optional arguments, and should obtain the active table 
from \texttt{self.\_context\_table}. However, the generated code changes the method signature by introducing an additional required \texttt{table} argument. Although the method body is superficially reasonable, the new signature violates the existing class contract and breaks downstream callers. This example shows how class-level generation can fail even when the generated code appears locally executable.

Example~3 presents a \textbf{field reconciliation failure} that arises during compositional generation. The merged class initializes the session identifier as \texttt{self.\_session\_id}, but the generated method later accesses \texttt{self.session\_id}. This small naming mismatch causes an \texttt{AttributeError} and indicates that the model preserves local method logic without consistently reconciling the shared state schema across the composed class. Such failures are particularly important in ClassEval-Pro because many tasks require multiple methods to operate over a shared internal representation.

Example~4 shows another form of \textbf{Logic Error}, where the generated code raises the wrong exception type. The required behaviour is to preserve standard dictionary semantics and raise \texttt{KeyError} for unknown strategies. Instead, the implementation introduces defensive handling with \texttt{ValueError}. Although this modification may look reasonable in isolation, it violates the expected interface behaviour and fails tests that depend on the original exception contract. This example highlights that correct class-level generation requires preserving not only return values, but also error semantics.

Example~5 demonstrates an \textbf{Integration Error} involving callback state handoff. The generated callback logic uses an \texttt{mmap} object and passes it to the callback, whereas the expected behaviour is to pass the original buffered stream payload consistently across \texttt{read()} and \texttt{\_trigger\_callback()}. The failure is therefore not a simple syntax or local logic problem, but a mismatch in the data object handed across method boundaries. This case illustrates how class-level tasks expose interaction bugs that only appear when multiple methods are executed together.

Example~6 shows a \textbf{Dependency Error} caused by a missing import. The generated method uses \texttt{os.path.exists} for path validation, but the corresponding \texttt{os} import is absent in the merged class. The local logic is otherwise reasonable, yet execution fails with a \texttt{NameError}. This example reflects a common dependency-management issue in compositional generation: models may introduce library-level dependencies during method synthesis without ensuring that the final class-level artifact contains the necessary imports.

Overall, these examples reinforce the need for a class-level benchmark that evaluates more than isolated method correctness. The failures are diverse, but they share a common theme: generated code often satisfies a local implementation intuition while violating global class consistency. ClassEval-Pro therefore exposes bottlenecks that are difficult to observe in function-level benchmarks, including signature preservation, shared-state reconciliation, callback contracts, exception compatibility, and import dependency management.

\section{Threats to Validity}

\textbf{External validity.} Our study targets Python exclusively; results may not generalize to statically-typed or systems-level languages. We mitigate temporal bias by sourcing only GitHub data from January 2025 onward, and our pipeline is language-agnostic by design.
\textbf{Internal validity.} LLM-assisted construction may introduce biases; we mitigate this via multi-stage quality control including compilation checks and test-suite validation with $>$90\% line coverage. Annotation subjectivity is addressed by three independent annotators with inter-annotator agreement (Kappa $>$ 0.7) and consensus-based conflict resolution.

\section{Conclusion}

We presented ClassEval-Pro, a benchmark targeting \textit{compositional code creation}---the underserved evaluation dimension between function-level synthesis and repository-level code modification. Our automated, contamination-resistant pipeline constructs 300 cross-domain class-level tasks spanning 11 domains from post-January-2025 GitHub code, replacing manual curation that limits scalability and temporal validity in prior benchmarks. Experiments with five frontier LLMs and five generation strategies reveal that compositional creation remains far from solved: the best model (Qwen3-480B) achieves only 45.6\% class-level Pass@1, with a 17.7-point gap separating the strongest and weakest models. Strategy choice substantially impacts performance---top-down and bottom-up approaches improve weaker models by up to 9.4 points, while compositional generation collapses to as low as 1.3\%. Error analysis over 500 manually annotated failed cases shows that logic errors (56.2\%) and dependency errors (38.0\%) dominate, identifying cross-method coordination as the core bottleneck. We release the benchmark, pipeline, and evaluation scripts to support future research.

\section*{Data Availability}\label{data_ava}

Our code and benchmark are available at \url{https://github.com/ian-Kappa/ClassEval-Pro}.

\section*{Acknowledgments}
This paper is supported by the National Key Research and Development Program of China (Grant No. 2023YFB4503802) and the Natural Science Foundation of Shanghai (Grant No. 25ZR1401175).

\bibliography{references} 
\bibliographystyle{ACM-Reference-Format}

\end{document}